\documentclass[twocolumn,
aps,prd,amsmath,showpacs,amssymb,superscriptaddress,nofootinbib,
longbibliography,eqsecnum,preprintnumbers]{revtex4-1}

\usepackage{amsmath, amsfonts, amsthm, amssymb, graphicx, color,hyperref}
\usepackage{subfigure}
\usepackage{tikz}
\usepackage{physics}
\usepackage{verbatim}
\usepackage{float}
\allowdisplaybreaks[3]

\def\({\left(}
\def\){\right)}
\def\[{\left[}
\def\]{\right]}
\def\d{\mathrm{d}}

\newcommand{\f}[2]{\frac{#1}{#2}}
\def \bal#1\eal  {\begin{align} #1 \end{align}}
\newcommand{\eref}[1]{{Eq.~(\ref{#1})}}
\newcommand{\be} {\begin{equation}}
\newcommand{\ee} {\end{equation}}
\newcommand{\bc}{\begin{center}}
\newcommand{\ec}{\end{center}}
\newcommand{\bim} {\begin{itemize}[noitemsep]}

\newcommand{\eim} {\end{itemize}}

\newcommand{\nn} {\nonumber\\}

\newcommand{\pd} {\partial}
\newcommand{\mc} {\mathcal}

\newcommand{\bi}{{\beta}}

\newcommand{\li}{{\lambda}}

\newcommand{\oi}{\omega}

\def\ba{\begin{eqnarray}}
\def\ea{\end{eqnarray}}

\newcommand{\mI}{\mathcal{I}}

\newcommand{\oneloop}{\tikz[c/.style={insert path={circle[radius=2pt]}}]
    \draw (0,0) ellipse (0.3 and 0.3) (0,-0.3)[c] ;}
\newcommand{\twoloop}{\tikz[c/.style={insert path={circle[radius=2pt]}}]
    \draw (0,0) ellipse (0.3 and 0.3) (-0.3,0)[c] (0.3,0)[c];}
\newcommand{\threeloop}{\tikz[c/.style={insert path={circle[radius=2pt]}}]
    \draw (0,0) ellipse (0.3 and 0.3) (0,-0.3)[c] (-0.26,0.15)[c] (0.26,0.15)[c];}
\newcommand{\floop}{\tikz[c/.style={insert path={circle[radius=2pt]}}]
    \draw (0,0) ellipse (0.3 and 0.3) (-0.3,0)[c] (0.3,0)[c] (0,-0.3)[c] (0,0.3)[c];}
\newcommand{\propagaheng}{\tikz[c/.style={insert path={circle[radius=2pt]}}]
    \draw (-0.3,0)--(0,0)[c]--(0.3,0) ; }
\newcommand{\propagatwo}{\tikz[c/.style={insert path={circle[radius=2pt]}}]
    \draw (-0.3,0)--(0,0)[c]--(0.3,0)[c]--(0.6,0);}
\newcommand{\propagathree}{\tikz[c/.style={insert path={circle[radius=2pt]}}]
    \draw (-0.3,0)--(0,0)[c]--(0.3,0)[c]--(0.6,0)[c]--(0.9,0);}
\newcommand{\propagafour}{\tikz[c/.style={insert path={circle[radius=2pt]}}]
    \draw (-0.3,0)--(0,0)[c]--(0.3,0)[c]--(0.6,0)[c]--(0.9,0)[c]--(1.2,0);}
\usetikzlibrary{shapes,shadows}
\tikzstyle{abstractbox} = [draw=black, fill=white, rectangle,
inner sep=10pt, style=rounded corners, drop shadow={fill=black,
    opacity=1}]
\tikzstyle{abstracttitle} =[fill=white]
\newsavebox{\myabstractbox}
\providecommand{\abstractnode}[2]{\begin{tikzpicture}%
    \node [abstractbox, fill=#1] (box)%
    {#2};%
    \node[abstracttitle, right=10pt] at (box.north west) {Step 1};
    \end{tikzpicture}}

\newcommand{\onedotlinetwo}{\tikz{\draw[fill = gray](0.3,0) circle (2pt);\draw [fill = white](0,0) circle (2pt);\draw(-0.3,0) .. controls (-0.1,0) .. (0,0) .. controls (0.3,0) .. (0.6,0) }}
\newcommand{\twodotlinetwo}{\tikz[c/.style={insert path={circle[radius=2pt]}}]
    \draw [fill=gray] (-0.3,0)--(0,0)[c]--(0.3,0)[c]--(0.6,0);}

\newcommand{\onedotlooptwo}{\tikz{\draw[fill = gray](0.3,0) circle (2pt);\draw [fill = white](-0.3,0) circle (2pt);\draw (0,0) ellipse (0.3 and 0.3) }}
\newcommand{\twodotlooptwo}{\tikz{\draw[fill = gray](0.3,0) circle (2pt);\draw [fill = gray](-0.3,0) circle (2pt);\draw (0,0) ellipse (0.3 and 0.3) }}

\newcommand{\onedotloopone}{\tikz{\draw [fill = gray](0,-0.3) circle (2pt);\draw (0,0) ellipse (0.3 and 0.3) }}

\newcommand{\onedotline}{\tikz[c/.style={insert path={circle[radius=2pt]}}]
    \draw[fill=gray] (-0.3,0)--(0,0)[c]--(0.3,0) ; }
\newcommand{\oneline}{\tikz[c/.style={insert path={circle[radius=2pt]}}]
    \draw (-0.3,0)--(0,0)[c]--(0.3,0) ; }
\newcommand{\twoline}{\tikz[c/.style={insert path={circle[radius=2pt]}}]
    \draw (-0.3,0)--(0,0)[c]--(0.3,0)[c]--(0.6,0);}

\begin{document}

\hfill {\footnotesize USTC-ICTS-19-21}

    \title{The Fredholm Approach to Charactize Gravitational Wave Echoes}

\author{Yu-Xin Huang}
\email{mocha@mail.ustc.edu.cn}
\affiliation{School of Physical Sciences, University of Science and Technology of China, Hefei, Anhui 230026, China}

\author{Jiu-Ci Xu}
\email{xjc61529@mail.ustc.edu.cn}
\affiliation{School of Physical Sciences, University of Science and Technology of China, Hefei, Anhui 230026, China}

\author{Shuang-Yong Zhou}
\email{zhoushy@ustc.edu.cn}
\affiliation{Interdisciplinary Center for Theoretical Study, University of Science and Technology of China, Hefei, Anhui 230026, China}

    \date{\today}

    \begin{abstract}
  
Gravitational waves are a sensitive probe into the structure of compact astronomical objects and the nature of gravity in the strong regime.  Modifications of near-horizon physics can imprint on the late time ringdown waveform, leaving behind a train of echoes, from which useful information about new physics in the strong gravity regime can be extracted. We propose a novel approach to compute the ringdown waveform and characterize individual echoes perturbatively using the Fredholm determinants, which can be intuitively represented via a diagrammatical scheme. Direct non-perturbative treatments can also be easily implemented for some cases. Numerically, the method is also effective and accurate even for relatively low resources.  
    
    \end{abstract}
    
    \maketitle
    \tableofcontents

    \section{Introduction}
    
The recent detections of gravitational waves (GWs) generated from some of the most extreme astronomical processes \cite{Abbott:2016blz, TheLIGOScientific:2017qsa} have opened up a new window to probe the strong gravity regime and extreme astrophysical environments, providing fresh opportunities to test the fundamental laws of physics at the deepest levels \cite{Yunes:2016jcc}.

Black holes are the most profound prediction of General Relativity (GR), which can be formed as the end states of gravitational collapses or other extreme processes. They are ideal playgrounds to theoretically test the principles and theories of quantum gravity,  and also have played important roles in model astronomy even before the detections of GWs. The defining feature of a black hole is the existence of the event horizon, which is one-way membrane for all matter sources and thought to be of paramount importance to understand the fundamental properties of quantum gravity. However, we can not directly observe the event horizon itself due to its very nature, not even in the recent GW events \cite{Cardoso:2017njb,Abramowicz:2002vt}.  Indeed, the exact physical nature of the event horizon is still being heavily debated: When quantum corrections are included, it has been suggested that at presumably the Planck distance away from the event horizon there may exist a firewall that burns any infalling observer, contrary to the classical picture that the observer does not encounter anything extraordinary when passing through the horizon \cite{Almheiri:2012rt}. Whilst for a classical black hole the horizon in the tortoise coordinate is at negative infinity, in the firewall proposal the black hole has a surface structure immediately outside the would-be event horizon at some finite tortoise coordinate associated with the Planck length. 
Similarly, if there exist horizonless exotic compact objects (ECOs) whose spacetimes resemble those of black holes outside some small distances away from the would-be the event horizons, the GWs generated from the corresponding processes of those ECOs will also fit the current observational data \cite{Cardoso:2017cqb, Cardoso:2017njb}. Various ECO models have been proposed such as boson stars \cite{Schunck:2003kk,Wheeler:1955zz}, fuzzballs \cite{Mathur:2005zp}, gravastars \cite{Mazur:2004fk}, and many others \cite{Holdom:2016nek, Barcelo:2010vc, Barcelo:2014cla}. 
A firewall or an ECO surface outside the would-be horizon can effectively introduces a barrier in the effective potential of the metric perturbation equation or a reflective boundary at a finite tortoise coordinate \cite{Cardoso:2017njb}. GWs can be trapped between the photosphere potential barrier and the reflective boundary, bounce back and forth within the trap and then transmit and propagate to the observer far away. This leads to a sequence of secondary pulses, dubbed {\it echoes}, after the initial ringdown pulse \cite{Cardoso:2016rao,  Cardoso:2017njb, Cardoso:2016oxy}. 

On the other hand, if gravity is modified in the strong field regime, which, despite precise GR tests in the solar system and other weak field environments \cite{Will:2014kxa}, has not been well constrained, extra features may arise in the effective potential of the perturbations around the black hole. For example, in mass-varying massive gravity \cite{Huang:2012pe}, because of the large enhancement of the graviton mass near the black hole horizon, the graviton potential can induce an extra potential barrier to the left of the photosphere potential peak, and thus GWs can also be trapped in between, leading to echoes in the late time ringdown waveform \cite{Zhang:2017jze}. Wormhole scenarios can also produce effective potentials with extra barriers \cite{Cardoso:2016rao, Wang:2018mlp}. 

Therefore, GW echoes might be a sensitive probe of near-horizon new physics, and recently there have been a lot of interests in studying the phenomenological and theoretical aspects of GW echoes \cite{Barcelo:2017lnx, Price:2017cjr, Nakano:2017fvh, Maselli:2017tfq, Volkel:2017ofl, Bueno:2017hyj, Volkel:2017kfj, Mark:2017dnq, Wang:2018gin, Pani:2018flj, Oshita:2018fqu,  Mannarelli:2018pjb, Tsang:2018uie, Testa:2018bzd, Correia:2018apm, Burgess:2018pmm, Urbano:2018nrs, Nielsen:2018lkf,  Lo:2018sep, Wang:2018mlp, Wang:2018cum, Konoplya:2018yrp, Conklin:2017lwb, Wang:2019szm, Li:2019kwa, Ren:2019afg, Ghersi:2019trn, Cardoso:2019apo, Uchikata:2019frs, Saraswat:2019npa, Tsang:2019zra,  Maggio:2019zyv}. Based some simple templates, it has been claimed that evidence of the existence of echoes is already contained in the current GW data \cite{Abedi:2016hgu, Abedi:2017isz}. However, for the current data, the statistical significance of the evidence is not sufficiently high to be a convincing detection \cite{Ashton:2016xff, Westerweck:2017hus}. In \cite{Price:2017cjr}, it is pointed out that for echoes coming from spacetime reflection conditions, the later ones are not simple repetition of the former in the sequence, which suggests that more sophisticated templates are generally desirable to accurately extract the echo signals. In \cite{Wang:2018mlp, Wang:2019szm, Li:2019kwa}, generic templates of non-equal interval echoes and possible scenarios that lead to un-equal intervals are initiated so as to accommodate more generic classes of ECOs or exotic scenarios to maximize possible information extraction in the GW data analysis. 

In \cite{Mark:2017dnq}, near-horizon exotic features are parametrized by a frequency dependent reflectivity at a fixed radius, and the full Green's function is then split into the Green's function of the corresponding black hole plus an extra term that can be expanded as a power series of the product of the reflectivities at the photosphere and the boundary and the phase delay between them. Consequently, the waveform observed far way can be re-processed as a power series of this product convoluted with the frequency domain wave amplitude at the would-be horizon. In this way, different terms in this series generates different individual echoes, which formalizes the physical picture of waves bouncing back and forth in the cavity formed by the photosphere barrier and the reflective boundary. Ref \cite{Correia:2018apm} proposed another another way to re-process the individual echoes perturbatively using the Dyson series. In this approach, the Green's function for the free wave equation (that is, without any potential) with the reflectively boundary is utilized instead, and the wave equation is solved with the Dyson series, a series of multi-dimensional integrals of the product of the Green's functions and the potentials. The individual echoes can be obtained via further expansions in terms of the reflectivity of the boundary. 

In this paper, we propose yet another method to efficiently compute the ringdown waveforms and intuitively re-process the individual echoes. Our method is based the Fredholm approach to solve the wave equation with the Fredholm determinants. An intuitive diagrammatical scheme can be utilized to evaluate two kinds of the Fredholm determinants. One kind of them are obtained by constructing all possible loop diagrams connecting all the vertices, and the other kind are obtained by cutting one of the propagators in each of the loop diagrams. Introducing a new kind of vertex, individual echoes can be re-processed as diagrams containing different numbers of the new vertices. For many cases, non-perturbative solutions can also be straightforwardly obtained without the need of re-summations of the perturbative series. Numerically, waveforms can be obtained in a simple yet efficient manner.

The paper is organized as follows. In Section~\ref{sec:Fredholm}, we will introduce the Fredholm formalism to compute the waveforms, perturbatively expand the solution with the Fredholm determinants, prove the convergence of the perturbative expansion and put estimates on the errors. In Section~\ref{sec:echo}, we will show how to non-perturbatively obtain the re-summed echo waveforms for a couple of toy models and how to extract individual echoes perturbatively. In Section~\ref{feynman}, we will introduce our diagrammatical scheme which can be used to compute generic waveform or isolate individual echoes. In Section~\ref{numerical}, we will numerically compute the waveforms in this formalism, check its reliability and show its efficiency.

    \section{Fredholm approach of ringdown waveforms}
    \label{sec:Fredholm}

In the ringdown phase, the metric perturbations for a spherical compact star can be expanded with spherical harmonics $Y_{lm}(\theta,\phi)$, and for a given $l$ mode the linear perturbation $\Phi(t,x)$ satisfies a wave equation of the form
\begin{equation}
\label{eq0}
    -\pdv[2]{\Phi(t,x)}{t}+\pdv[2]{\Phi(t,x)}{x}  - V(x)\Phi(t,x)=0 ,
\end{equation}
where the effective potential $V(x)$ depends on the background metric, the spherical harmonic number $l$ and the spin of the perturbation mode. $x$ is the tortoise (radial) coordinate that makes the background metric conformally flat for the temporal and radial coordinates. In order to obtain the ringdown waveform, the following asymptotic behavior is imposed at spatial infinity
\begin{equation}
   \Phi(t,x)\sim e^{i\omega(x-t)} , \quad x\rightarrow\infty  .
\end{equation}
On the left hand side, to capture various boundary conditions of potential ECOs, we impose the following boundary condition
\begin{equation}
    \Phi(t,x)\sim e^{-i \omega(x+t)}+R(\omega) e^{i \omega(x-t)}, ~~~x=-L  ,
\end{equation}
where $R(\omega)$ is the frequency dependent reflectivity and $L$ is to the left of the photosphere potential barrier. When the left hand side is also open ($x\to-\infty$), similar to the case of the Schwarzschild black hole, we may set $R(\omega)=0$ and $L\to \infty$. Equivalently, we may represent the effects of the extra potential barriers with some specific $R(\omega)$ at $x=-L$.

After Laplacian transform 
\begin{equation}
\phi(\omega,x)=\int_0^{\infty} \Phi(t,x)e^{i\omega t}\dd t  ,
\end{equation}
the wave equation in frequency domain is given by
    \begin{equation}
    \label{phiequ1}
   \phi''(\omega,x)+(\omega^2-V(x))\phi(\omega, x) =\mathcal{I}(\omega,x)   ,
    \end{equation}
where ${}'$ stands for a derivative with respect to $x$ and 
\begin{equation}
\mathcal{I}(\omega,x)= i\omega \Phi(0,x)-\pd_t \Phi{}(0,x). 
\end{equation}
 In the following we will often suppress the $\omega$ dependence  $\phi(x)\equiv \phi(\omega,x)$. To get the solution in the time domain, we can solve Eq.~(\ref{phiequ1}) for different frequencies and perform the inverse Laplacian transform.

\subsection{The Fredholm integral equation}
\label{setting}

In this section, we shall solve Eq.~(\ref{phiequ1}) by the Fredholm method. 
Suppose that $G(x,y)$ is the Green's function of
\begin{equation}\label{green0}
 \(\dv[2]{x}+\omega^2 \)  G(x,y) =\delta(x-y)
\end{equation}
with appropriate boundary conditions. Then \eref{phiequ1} can be cast as the Fredholm integral equation,

    \begin{equation}
    \label{fredhomeq}
    f(x)=\phi(x)-\lambda\int\dd y K(x,y)\phi(y)   ,
    \end{equation}
where 
\begin{equation}\label{f(x)}
f(x)=\int \dd y G(x,y) \mc{I}(y)   
\end{equation}
and the Kernel function is
\begin{equation}
K(x,y)=G(x,y) V(y)   .
\end{equation}
For later convenience, an explicit order parameter $\lambda$ has been introduced to count the expansion order in terms of interaction $V(y)$, which may be set to 1 in the end results. Assuming that $\{\alpha_i(x)\}$ and $\{\beta_i(y)\}$ are appropriate sets of base functions for radial variables $x$ and $y$ respectively, the kernel function can be expanded  as
    \begin{equation}
    \label{Kdef1}
    K(x,y)=\sum_{j=1}^\infty \alpha_j(x)\beta_j(y)   .
    \end{equation}
With this expansion, \eref{fredhomeq} can be then written as a system of linear algebraic equations
    \begin{equation}
    \label{YXlin}
    Y_{i}=X_{i}-\lambda \sum_{j} B_{ij}X_{j}   ,
    \end{equation}
where we have defined
\begin{equation}
\begin{split}
X_i &=\int \d x \beta_i(x) \phi(x), \quad Y_i = \int \d x \bi_i(x) f(x),\\
B_{ij} &= \int \d x \bi_i(x) \alpha_j(x)   . 
\end{split}
\label{Bdef1}
\end{equation}
The solution of \eref{YXlin} is given by
    \begin{equation}
    X_{i}=\frac{{\Delta_{i}}(\lambda)}{{\Delta}(\lambda)} ,
    \end{equation}
where $\Delta(\lambda)=\det(I-\lambda B)$, $B=(B_{ij})$ and $\Delta_{i}(\lambda)$ is almost ${\Delta}(\lambda)$ but with the $i$-th column of $I-\lambda B$ replaced by  $Y=(Y_{1},Y_{2},\dots)^T$. Plugging the solution back into \eref{fredhomeq}, we find that
    \begin{equation}
    \label{Delta}
    \phi(x)=f(x)+ \f{\lambda}{\Delta(\lambda)}\sum_{j} \alpha_j(x) \Delta_j(\lambda)  .
    \end{equation}
Notice that $\sum_{j} \alpha_j(x) \Delta_j(\lambda) $ can  be rewritten as
\begin{align}
&~~~~ \sum_{j} \alpha_j(x) \Delta_j(\lambda)  
\\
&= \sum_{j}\alpha_{j}(x)\left|
\begin{matrix}
c_{11} & \cdots & Y_{1} & c_{1,j+1}& \cdots \\
c_{21} & \cdots & Y_{2} & c_{2,j+1} &\cdots \\
 &     &   \cdots  &      & \vdots  
\end{matrix}
\right| \\
&=
-\sum_{j}(-1)^{j} \alpha_{j}(x)\left|
\begin{matrix}
Y_{1} & c_{11} & \cdots   & \hat{c}_{1j} & \cdots  \\
Y_{2} &c_{21}  & \cdots  & \hat{c}_{2j} & \cdots \\
 &     &   \cdots  &      & \vdots  
\end{matrix}
\right|
\\
&=
\left|
\begin{matrix}
0     & \alpha_{1}(x) & \alpha_{2}(x) & \cdots \\
Y_{1} & c_{11} & {c_{12}} & \cdots  \\
Y_{2} & c_{21} & {c_{22}} & \cdots \\
 &        \cdots  &     &  \vdots  
\end{matrix}
\right|   ,
\end{align}
where $c_{ij}$ are the matrix elements of $I-\lambda B$ and $\hat{c}_{1j}$, $\hat{c}_{2j}$, etc.~denote missing elements in the corresponding places of the determinant.
Therefore, the solution in the frequency domain is given by
\begin{equation}
\phi(x) =f(x)-\lambda \int \d y f(y)\frac{\Delta_K(x,y,\lambda)}{\Delta(\lambda)} ,
\end{equation}
where 
\begin{equation}
\Delta_{K}(x,y,\lambda)=\left|
\begin{matrix}
0     & \alpha_{1}(x) & \alpha_{2}(x) & \cdots \\
\beta_{1}(y) & c_{11} & {c_{12}} & \cdots  \\
\beta_{2}(y) & c_{21} & {c_{22}} & \cdots \\
 &        \cdots  &    &   \vdots  
\end{matrix}  
\right|  .
\end{equation}
Then the solution in the time domain can be written as
    \be
    \label{invLap}
   \Phi(t,x)  =\int_{-\infty+\beta i}^{\infty+\beta i}\phi(\omega,x)e^{-i\omega t}\frac{\dd \omega}{2\pi}  ,
   \ee
where $\beta$ is chosen such that all the singularities of $\phi(\omega,x)$ are below the integration path in the complex $\oi$ plane.

\subsection{Examples} 
\label{examples}

Let us do a couple of examples to demonstrate how this formalism works. A trivial example would be when $K(x,y)=xy$ within the interval $y\in [0,1]$ and $K(x,y)=0$ otherwise.
\begin{equation} 
\label{trivialexa}
\phi(x)=f(x)+x\int_{0}^{1} y\phi(y) \dd y   ,
\end{equation}
where we have set $\lambda=1$. For this case, the solution is simply
\begin{align}
\phi(x)
&=f(x)-\int_{0}^{1}\frac{\begin{vmatrix}
0 & x \\
y & 2/3  \end{vmatrix}}{\begin{vmatrix}
2/3
\end{vmatrix}}f(y)\dd y
\\
&=f(x)+\frac{3}{2}x \int_{0}^{1} yf(y) \dd y  . \label{trivial}
\end{align}

Next, let us consider a system with a Schwarzschild-like boundary condition, for which case the free Green's function is
\be
G(x,y)=\frac{e^{i\omega|x-y|}}{2i\omega}  .
\ee
We seek the asymptotical wave solution observed far away, that is, we want to evaluate the waveforms at large and positive $x$, where the observer is located, so we have $|x-y|=x-y$. Then the kernel in this case is manifestly separable, we can set $K(x,y)=\alpha(x)\beta(y)$ with
\begin{equation}
\alpha(x)=\frac{e^{i\omega x}}{2i\omega} , \quad\beta(y)=e^{-i\omega y}V(y)  .
\end{equation}
Thus the far-field solution is not much different from the trivial case above,  and we can get
\begin{equation}
\phi(x)=f(x)+\!\int\!\!\dd x_1\frac{e^{i\omega(x-x_1)}}{2i\omega} \frac{V(x_1)f(x_1)}{1-\frac{1}{2i\omega}\int\!\!\dd y V(y)}  .
\end{equation}
To compare with the example used in \cite{Correia:2018apm}, let us restrict to the Dirac delta potential $V(x_1)=2V_{0}\delta(x_1-y)$, with $V_0$ being a positive constant, and we have
\begin{equation}
 \phi(x) = f(x)+\int\frac{G(x,x_{1})V(x_{1})f(x_1)}{1-V_{0}/i\omega}  \dd x_{1} ,
\end{equation}
and the pole at $\omega =-iV_0$ gives the quasi-normal mode of the system. This is the same as the re-summed Dyson series (see \cite{Correia:2018apm})

Therefore, for cases where the kernel function $K(x,y)$ can be separated with a sum of finite terms, the Fredholm approach can directly give the nonperturbative solution. We want to emphasize that \eref{Kdef1} with an infinite summation can always be achieved, but an exact separation with a finite summation is not always possible. However, numerically, the kernel can be always approximated with a finite separable sum for a given precision, as we will see shortly.\\

\subsection{Perturbation scheme}

We can also obtain perturbative solutions in the Fredholm approach by expanding both $\Delta_K(x,y,\lambda)$ and ${\Delta}(\lambda)$ in terms of $\lambda$, and  we have
\bal
\phi(x)&=f(x)-\lambda \int \d y f(y)\frac{\Delta_K(x,y,\lambda)}{\Delta(\lambda)}
\nn 
\label{fredsol}
&=f(x)+\lambda\int \dd y f(y)\frac{\sum_{n=0}^{\infty}\frac{(-\lambda)^{n}}{n!}A_{n}(x,y)}{\sum_{n=0}^{\infty}\frac{(-\lambda)^{n}}{n!}D_{n}}    ,
\eal
where the Fredholm determiants $A_{n}(x,y)$ and $D_{n}$ are given by
\begin{widetext}
\begin{equation}\label{Andef}
\begin{split} 
&A_{n}(x,y)=\int{\prod_{i=1}^{n}\dd x_{i}}
\left|\begin{matrix}
K(x,y) & K(x,x_{1}) & K(x,x_{2}) & \cdots & K(x, x_{n})\\
K(x_{1},y) & K(x_{1},x_{1}) & K(x_{1},x_{2}) & \cdots & K(x_{1},x_{n})\\
& \cdots& &\vdots &\\
K(x_{n},y) & K(x_{n},x_{1}) & K(x_{n},x_{2}) & \cdots & K(x_{n},x_{n})
\end{matrix}\right|   ,
\end{split}
\end{equation}
\begin{equation}\label{Dndef}
\begin{split}
&D_{n}=\int{\prod_{i=1}^{n}\dd x_{i}}
\left|\begin{matrix}
K(x_{1},x_{1}) & K(x_{1},x_{2}) & K(x_{1},x_{3}) & \cdots & K(x_{1}, x_{n})\\
K(x_{2},x_{1}) & K(x_{2},x_{2}) & K(x_{2},x_{3}) & \cdots & K(x_{2},x_{n})\\
& \cdots& &\vdots &\\
K(x_{n},x_{1}) & K(x_{n},x_{2}) & K(x_{n},x_{3}) & \cdots & K(x_{n},x_{n})
\end{matrix}\right|  .
\end{split}
\end{equation}
\end{widetext}
We will show why the expansion of $\Delta_K(x,y,\lambda)$ and ${\Delta}(\lambda)$ can be achieved in terms of $A_{n}(x,y)$ and $D_{n}$ respectively in Section \ref{feynman} via a diagrammatical scheme. In doing so, one obtains a Pade series in terms of $\lambda$, which is a double Taylor expansion both in the numerator and the denominator and whose coefficients are related to the eigen-polynomials of transition matrix $B$.  As we shall later, if the kernel contains some exotic features near the would-be horizon such as a ``mirror'' with a frequency dependent reflectivity $R(\omega)$, one can also get a power series in $R(\omega)$ by re-processing the $\lambda$ Pade series. Thus, we have reduced the problem of solving the differential equation \eref{phiequ1} to evaluating the determinants and integrations of the kernel $K(x,y)$. As we shall see later, these determinants can be conveniently computed with the diagrammatical method outlined in Section \ref{feynman}.

\subsection{Convergence and error estimate}
\label{appA}

To numerically calculate the waveforms to a desired accuracy, we can choose a sufficiently large interval $[a,b]$ for the integration in the Fredholm integral equation and truncate the Pade expansion to a sufficiently large order $N$. In the following, we shall prove the convergence of the Pade series and then estimate of the errors arising from the different sources of approximation.

First, note that \eref{Andef} and \eref{Dndef} allow us to use the Hadamard inequality to find the upper bounds:
\bal
|A_{n}(x,y)|&< \sqrt{(n+1)^{n+1}}\tilde M^{n+1}(b-a)^{n}  ,\\
|D_{n}|&< \sqrt{n^{n}}\tilde M^{n}(b-a)^{n}   ,
\eal
where $\tilde M$ is the upper bound of $|K_{N}(x,y)|$. So the numerator of Eq.(\ref{fredsol}) is bounded by
\begin{equation}
\left|\sum_{n}\frac{(-\lambda)^{n}}{n!}A_{n}\right|< \sum_{n=0}^{\infty}\sqrt{(n+1)^{n+1}}\tilde M^{n+1}(b-a)^{n} \frac{\lambda^{n}}{n!}     .
\end{equation}
Since $\frac{z^{n}}{n!}< e^{z}$  for positive number $z$, we get 
\bal \label{fangsuo}
&~~~~~\f{\sqrt{(n+1)^{n+1}}}{n!}\tilde M^{n+1}(b-a)^{n} \lambda^{n} 
\nn
&< \frac{e^{n+1/2} \tilde M^{n+1} (b-a)^{n}(n+1)^{1/2}\lambda^{n} }{\sqrt{n!}}   .
\eal
Since $\lim_{n\rightarrow\infty} (n!)^{1/n} /(n+1)^{1/2}=\infty$, we see that when $\Delta_K(x,y,\lambda)$ is expanded in terms of $A_n(x,y)$, the convergence radius for $\lambda$ is infinite. Of course, a similar argument holds for $\Delta(\lambda)$.  This means that the solution we have constructed in \eref{fredsol} is convergent, and the limit of $N\to\infty$ leads to the exact solution.

Setting $\lambda=1$, our approximate solution is given by
\begin{equation}
\phi_{\rm app}(x)=f(x)+\!\int_{a}^{b}\!\frac{\sum_{n=0}^{N}\frac{(-\lambda)^{n}}{n!}A_{n}(x,y)}{\sum_{n=0}^{N}\frac{(-\lambda)^{n}}{n!}D_{n}} f(y) \dd y  ,
\end{equation} 
and the numerical errors can be bounded the following three parts:
\bal
\label{errorapprox}
&~~~~\left|\phi(x)-\phi_{\rm app}(x)\right| 
\nn
&< \left|\int_{-\infty}^{a}\frac{\Delta_{K}(x,y)}{\Delta}f(y)\dd y\right|\nn
&~~~~~+\left|\int_{b}^{\infty}\frac{\Delta_{K}(x,y)}{\Delta} f(y)\dd y\right|\nn
&~~~~~+\left|\int_{a}^{b}\(\frac{\Delta_{K}(x,y)}{\Delta}-\frac{\Delta^{(N)}_{K}(x,y)}{\Delta^{(N)}}\)f(y)\dd y\right| ,
\eal
where $\Delta^{(N)}_{K}(x,y)$ and $\Delta^{(N)}$ denote the $N$-th terms of $\Delta_{K}(x,y)$ and $\Delta$ respectively.

To estimate the errors introduced by truncating the $\li$ expansion to order $N$, we note that  \eref{fangsuo} implies that $|\Delta_{K}(x,y)-\Delta^{(N)}_{K}(x,y)|$ is bounded by
\bal
&~~~~|\Delta_{K}(x,y)-\Delta^{(N)}_{K}(x,y)|=\left|\sum_{n=N+1}^{\infty}\frac{(-1)^{n}}{n!}A_{n}(x,y)\right|
\nn
&<\frac{e^{n/2}\tilde M^{n+1}(b-a)^{n}\sqrt{(n+1)}}{\sqrt{n!}}\nn
&<2\sqrt{\frac{e^{N}(b-a)^{N+4}\tilde{M}^{2N+8}}{N!}}\sum_{n=N+1}^{\infty}\prod_{l=N+1}^{n}\sqrt{\frac{e(b-a)\tilde{M}^{2}}{l}}\nn
&<2\sqrt{\frac{e^{N}(b-a)^{N+4}\tilde{M}^{2N+8}}{N!}}\left({1-\frac{e(b-a)\tilde M^{2}}{N}} \right)^{-\f12}\nn
&\equiv \delta_{N}  .
\eal
Also, we can show that $|\Delta-\Delta^{(N)}|$ is bounded $\epsilon_{N}$ in a similar manner, particularly suppressed by $1/\sqrt{N!}$. Therefore, one finds
\bal
&~~~~\left|\int_{a}^{b}\(\frac{\Delta_{K}(x,y)}{\Delta}-\frac{\Delta^{(N)}_{K}(x,y)}{\Delta^{(N)}}\)f(y)\dd y\right| \\
&<\left|\int_{a}^{b}f(y)\dd y \frac{-\epsilon_{N}\Delta_{K}(x,y)+\delta_{N}\Delta}{\Delta^{2}}\right|+O(\epsilon^{2}_{N},\delta^{2}_{N})  ,\nonumber
\eal
which is suppressed by $1/\sqrt{N!}$.

To estimate the errors arising from numerical integrations in a finite interval, we need to get a handle of how integrations in the infinite intervals $(-\infty,a)$ and $(b,\infty)$ will contribute to $A_{n}$ and $D_{n}$. For concreteness, we shall do the estimate on the case of the Schwarzschild effective potential, but similar estimates apply for the cases of ECOs as long as the effective potential still falls off sufficiently fast away from the peak:
\bal
\label{Vgr}
V(y) &=\(1-\dfrac{1}{r(y)}\)\(\dfrac{l(l+1)}{r^{2}(y)}+\dfrac{1}{r^{3}(y)}\)  ,
\eal
where $l$ is the $l$ from the spherical harmonics, $r(y)$ is implicitly determined by $y =r(y)+\ln({r(y)-1})$, and in this subsection we have restricted to the scalar perturbation and chosen the units such that $2GM=1$ ($G$ being the Newton's gravitational constant and $M$ being the mass of the black hole).

Obviously, we need to choose the interval $[a,b]$ such that $V (y=a,b)$ are sufficiently small. When $r(y)$ is close to 1, which corresponds to $y$ being large and negative, we have
\begin{equation}
y=r+\ln(r-1)>\ln(r-1) ~~~ {\rm ie,} ~~~e^{y}>r-1  ,
\end{equation}
which leads to
\bal
     V(y) &<e^{y}(l^{2}+l+1)  ,\\
 \int_{-\infty}^{a}V(y)\dd y &<e^{a}(l^{2}+l+1)   .
\eal
On the other hand, when $r$ is large, which corresponds to the case when $y\to\infty$, we have
\begin{equation}
    \frac{y}{2}<r<y   ,
\end{equation}
which leads to
\begin{equation}
 \begin{split}
 V(y)&<\frac{4l(l+1)}{y^{2}}+\frac{8}{y^{3}} ,  \\
 \int_{b}^{+\infty}|V(y)|\dd y &<\frac{4l(l+1)}{b}+\frac{4}{b^{2}} .
 \end{split}
\end{equation}

Note also that the maximum value of $G(x,y)=\frac{e^{i\omega|x-y|}}{2i\omega}$ is just $w=\frac{1}{2|\omega|}$ and
\bal
A_{n}(x,y) & =\int\prod_{i=1}^{n}\dd x_{i} V(x_{1})\dots V(x_{n})V(y)
\nn
& ~~~~~~~~~~\cdot\left|
\begin{matrix}
G(x,y) &\dots & G(x,x_n)\\
\dots & \dots &\dots\\
G(x_{n},y) & \dots & G(x_{n},x_{n})  
\end{matrix}
\right| ,
\eal
which can be bounded by the Hadamard inequality.  Denoting $A^{[a,b]}_{n}(x,y)$ as $A_n(x,y)$ but with the integration limiting to the interval $[a,b]$, we have
\bal
\delta A_{n}(x,y)&= |A_{n}(x,y)-A^{[a,b]}_{n}(x,y)|\nn
&< V(y)\sqrt{(n+1)^{n+1}}w^{n+1}\nn
&~~~~~\cdot\left[\(\int_{-\infty}^{\infty}\dd x V(x)\)^{n}-\(\int_{a}^{b}V(x)\dd x\)^{n}\right]\nn
&< \frac{V(y)\sqrt{(n+1)^{n+1}}w^{n+1}\tilde{V}^{n}\epsilon(a,b)}{\tilde{V}-\epsilon(a,b)}\nn
&< \frac{2V(y)\epsilon(a,b)}{\tilde{V}^{2}}\tilde{m}^{n+1}\sqrt{(n+1)!}  ,
\eal
where we have defined 
\bal
 \tilde{V}&=\int_{a}^{b}V(x)\dd x  ,\\ 
\epsilon(a,b) &=e^{a}(l^{2}+l+1)+\frac{4l(l+1)}{b}+\frac{4}{b^{2}} , \\
 \tilde{m}&=e^{1/2}w\tilde{V} ,
\eal
and chosen $a$ and $b$ such that $\tilde{V}-\epsilon(a,b)>\frac{\tilde{V}}{2}$. 
So the finite interval error in the numerator is bounded by
\
\bal
 \label{errorestimatean}
&~~~~ \left|\sum_{n=0}^{\infty}\frac{(-1)^{n}}{n!}\delta A_{n}(x,y)\right|
\nn
&< \frac{2\epsilon(a,b)V(y)}{\tilde{V}^{2}}\sum_{n=0}^{\infty} \frac{\tilde{m}^{n+1} \sqrt{n+1} }{\sqrt{n!}} 
 \nn
&= \text{(first 3 terms)}+ \frac{2\epsilon(a,b)V(y)}{\tilde{V}^{2}}
\nn
&~~~~~~~~~~~~~~\cdot \sum_{n=3}^{\infty} \frac{\tilde{m}^{n+1}}{\sqrt{(n-3)!}}\sqrt{\frac{n+1}{n(n-1)(n-2)}}
\nn
&< \text{(first 3 terms)}+\frac{2\epsilon(a,b)V(y)}{\tilde{V}^{2}}
\nn
&~~~~~~~~~\cdot \left[\(\sum_{n=3}^{\infty}\frac{\tilde m^{2n+2}}{(n-3)!}\)\(\sum_{n=3}^{\infty}\frac{n+1}{n(n-1)(n-2)}\) \right]^{\frac{1}{2}}
\nn
&=\frac{2\epsilon(a,b)V(y)}{\tilde{V}^{2}}\tilde{f}(\tilde{m})  ,
\eal
where
$\tilde{f}(\tilde{m})\equiv \tilde{m}+\sqrt{2}\tilde{m}^{2}+\frac{\sqrt{6}}{2}\tilde{m}^{3}+\frac{\sqrt5}{2}\tilde{m}^{4}e^{\tilde{m}^{2}/2}$. For the denominator, similarly, we have
\bal
|\delta D_{n}|=|D^{\infty}_{n}-D^{(a,b)}_{n}|&<\tilde{m}^{n}\sqrt{n!}\frac{2\epsilon(a,b)}{\tilde{V}}\\
\left|\sum_{n=1}^{\infty}\frac{(-1)^{n}}{n!}\delta D_{n}\right|&< \frac{2\epsilon(a,b)}{\tilde{V}}(1+e^{\tilde{m}^{2}})\tilde{m}
\label{errorestimatedn}
\eal
Therefore, we see that the errors introduced by using the finite interval $[a,b]$ to compute $A_{n}(x,y)$ and $D_{n}$ is bounced by $\epsilon(a,b)$, which decreases exponentially with $a$ ($a$ being negative) but only inversely proportionally with $b$. This suggests that while a relatively small $|a|$ is sufficient, it is imperative to choose relatively large $b$.

Combining \eref{errorestimatean} and \eref{errorestimatedn}, we can get
\begin{equation}
\left|\frac{\Delta_{K}(x,y)}{\Delta}\right|< \frac{V(y)}{ \tilde{V}\Delta}\tilde{f}(\tilde{m})  .
\end{equation}
We see that the bound is independent of $a$ and $b$. Therefore, the first two terms in \eref{errorapprox} are bounded by
\bal
    \left| \int_{-\infty}^{a}\frac{\Delta_{K}(x,y)}{\Delta}f(y)\dd y\right|&+\left|\int_{b}^{\infty}\frac{\Delta_{K}(x,y)}{\Delta} f(y)\dd y\right|   \nn
   & < \frac{m_{2}\epsilon(a,b)}{\tilde{V}}f(\tilde{m})
\eal
where $m_{2}$ is the maximum value of $f(y)$ in $(-\infty,a)\cup(b,\infty)$. The bound is also proportional to $e^{a}/{b}$.

These are of course the upper bounds, and in actual calculations there may be cancellations, which will lead to better accuracies.

\section{Echoes in the Fredholm approach}
\label{sec:echo}

Echoes in the ringdown waveform are signatures of new physics in the strong gravity regime. They can arise if the effective potential is modified with a reflective mirror or an extra potential barrier to the left of the photosphere peak. In the following, we shall first demonstrate how to re-process the echoes in the Fredholm approach with two separable examples, which allows for direct nonpeturbative treatments, and then show how to extract echo signals systematically by perturbation theory.

\subsection{Delta potential with a mirror}

First, we consider a Dirac delta potential 
\be
V(x)=2V_{0}\delta(y)
\ee
with a reflective mirror at $x=-L$. In this case, using \eref{mirgf}, the integral kernel can be written as 
\begin{align}
    K(x,y)
    &=\frac{e^{i\omega |x-y|}}{2i\omega} 2V_{0}\delta(y)+R\frac{e^{i\omega(x+y)}}{2i\omega}  2V_{0}\delta({y})    .
\end{align}
Using the property of the delta function, the kernel can be written as
\begin{equation}
  K(x,y) = \sum_{i=1,2} \alpha_i(x) \beta_i(y)  ,
\end{equation}
with
\begin{align}
\alpha_{1}(x) &=\frac{e^{i\omega|x|}}{2i\omega}  , & \alpha_{2}(x)&=\frac{e^{i\omega x}} {2i\omega}  ,
\\
\beta_{1}(y)&=2V_{0}\delta(y)  , & \beta_{2}(y)&=R(\omega)2V_{0}\delta(y) .
\end{align}
Calculating the transition matrix and setting $\lambda=1$, it is straightforward to get
\begin{equation}
\label{exactdelta}
\phi(x)=f(x)+\frac{\frac{V_{0}}{i\omega}(e^{i\omega|x|}+R e^{i\omega x})}{1-\frac{V_{0}}  {i\omega}(1+R)}f(0) ,
\end{equation}
where $f(x)$ is simply the first term in the Born approximation
\begin{equation}
f(x)=\int\!\!\dd y \(\frac{e^{i\omega |x-y|}}{2i\omega} +R(\omega)\frac{e^{i\omega(x+y)}}{2i\omega} \) \mI (\omega,y)   .
\end{equation}
The denominator of \eref{exactdelta} can be expanded as
\begin{align}
\frac{1}{1-\frac{V_{0}}{i\omega}(1+R)}&=\sum_{n=0}^{\infty}\(\frac{V_{0}}{i\omega}\)^{n}\sum_{k=0}^{n}\binom{n}{k}R^{k}\\
&=\frac{i\omega}{V_0}\sum_{k=0}^{\infty}R^{k}\sum_{n=k}^{\infty} \binom{n}{k}\(\frac{V_{0}}{i\omega}\)^{n+1}\\
&=\frac{i\omega}{V_0}\sum_{k=0}^{\infty}R^{k}R_{\delta}^{k+1} \label{eq2.59}  ,
\end{align}
where $\binom{n}{k}$ is the binomial coefficients ``$n$ choose $k$" and we have defined reflectivity of the delta potential
\begin{equation}
R_{\delta}=\sum_{n=1}^{\infty} \(\frac{V_{0}}{i\omega} \)^{n}=-\frac{V_{0}}{V_{0}-i\omega}  .
\end{equation}
Substituting \eref{eq2.59} into \eref{exactdelta}, we get 
\begin{equation}
\phi(\omega,x)=\phi_{0}(\omega,x)+\sum_{n=1}^{\infty} R^{n}\phi_{n}(\omega,x)  ,
\end{equation}
which is a power series in terms of the reflectivity $R(\oi)$ of the mirror. The the $n$-th echo in the frequency domain is just the wave packet that has bounced $n$ time by the mirror, thus the term with the $n$-th power of $R(\oi)$ \cite{Mark:2017dnq}, and therefore reads
\begin{align}\label{simple}
 &~~~~R^n\phi_{n}(\omega,x)
 \nn
 &=\int\!\! \bigg(R_{\delta}^{n+1}e^{i\omega(|x|+|y|)}+R_{\delta}^{n-1}e^{i\omega(x+y)}+
\nn
&~~~~~~~~~~ R_{\delta}^{n}(e^{i\omega(|x|+y)}+e^{i\omega(x+|y|)}) \bigg) \frac{\mI(\omega,y)}{2i\omega}\dd y   ,
\end{align} 
for $n\geq 1$; $\phi_{0}$ is given by \eref{simple} with $R_{\delta}^{-1}$ setting to 0.
This is in complete agreement with \cite{Correia:2018apm}, which uses the Dyson series approach.

\subsection{Double delta potential}
\label{sec:ddp}

Now, we want to compute the echoes for the double delta potential
\be
V(x) = 2V_{0}\delta(y)+R(\omega) 2V_{0}\delta(y-P)  ,
\ee
 with an open boundary on the left hand side, that is, the left hand boundary is at $x=-L$ with $L\to \infty$. As we shall see momentarily, this is very much the same as the previous case in the Fredholm approach, while in the Dyson series approach it is more involved if one needs to get the full non-perturbative solution.

In this case, the kernel function is 
\begin{equation}
   K(x,y)=\frac{e^{i\omega|x-y|}}{2i\omega}[ 2V_{0}\delta(y)+R(\omega) 2V_{0}\delta(y-P)] ,
\end{equation}
which can be written as
\begin{equation}
  K(x,y) = \sum_{i=1,2} \alpha_i(x) \beta_i(y)  ,
\end{equation}
with
\bal
\alpha_{1}(x) &=\frac{V_{0}}{i\omega}e^{i\omega|x|}  , & \alpha_{2}(x)&=\frac{V_{0}}{i\omega}e^{i\omega|x-P|}  ,
\\
\beta_{1}(y)&=\delta(y)  ,  & \beta_{2}(y)&=R(\omega)\delta(y-P) .
\eal
Going through steps similar to the case of the delta potential with a mirror, we can get
\bal
&\phi(x)=f(x)\\
&~~+\int\dd y\frac{\frac{V_{0}}{i\omega}e^{i\omega|x-y|}(\delta(y)+R\delta(y-P))}{1-(1+R)\frac{V_{0}}{i\omega}+R(\frac{V_{0}}{i\omega})^{2}(1-e^{2i\omega|P|})} f(y)\nn
&~~+\int\dd y \frac{R(\frac{V_{0}}{i\omega})^{2}\left(e^{i\omega(|x-P|+|y-P|)}-e^{i\omega|x-y|}\right)\delta(y)}{1-(1+R)\frac{V_{0}}{i\omega}+R(\frac{V_{0}}{i\omega})^{2}(1-e^{2i\omega|P|})}f(y)\nn
&~~+\int\dd y \frac{R(\frac{V_{0}}{i\omega})^{2}\left(e^{i\omega|x|+|y|}-e^{i\omega(|x-y|)}\right)\delta(y-P)}{1-(1+R)\frac{V_{0}}{i\omega}+R(\frac{V_{0}}{i\omega})^{2}(1-e^{2i\omega|P|})}f(y)  ,
\label{ddpsol}
\eal
which is the full non-perturbative solution in terms of $R$. The $n$-th echo signal is simply the $n$-th term in the expansion of \eref{ddpsol} in terms of $R$. This is different from the Dyson series approach where one gets the perturbative solutions order by order, which seems to be difficult to be re-summed to a full non-perturbative form. 

Notice that the phase delay factor $Re^{2i\omega|P|}$ appears in denominators of the terms in \eref{ddpsol} and there are $R$ factors in numerators, so  when expanded in terms of $R$, the $R^{n}$ term will come with either a phase delay factor $e^{2ni\omega|P|}$ or $e^{(2n+1)i\omega|P|}$. This basically reflects the physical intuition that the $n$-th echo comes from waves that are reflected  $n$ or $n+1$ times between the two potential barriers.

\subsection{Perturbation scheme}

In the Fredholm formalism, we can also decompose the waveforms into different echo signals via a perturbative expansion. For the case where the boundary $x=-L$ is a reflective mirror. The Green's function in Eq.~\ref{green0} is modified to
\begin{equation} 
\label{mirgf}
G(x,y)=\frac{e^{i\omega|x-y|}}{2i\omega}+R(\omega)\frac{e^{i\omega(x+y)}}{2i\omega} .
\end{equation}
On the other hand, for the case where the potential can be split into two localized parts
\begin{equation}
 {V}(x)= \bar V(x)+R(\omega)V_{\rm Q}(x)   ,
\end{equation}
the kernel can be written as
\begin{equation}
  K(x,y)= \frac{e^{i\omega |x-y|}}{2i\omega}\bar V(y) +R(\omega)\frac{e^{i\omega|x-y|}}{2i\omega}V_Q(y)    ,
\end{equation}
where often $R(\omega)$ just has a trivial $\omega$ dependence. For either case, the kernel function can be split into two parts
\begin{equation}
   K(x,y) =\bar K(x,y)+R(\omega) Q(x,y) \label{reflection}   .
\end{equation}
Substituting this into the definition of $D_n$, we find
\begin{equation}
 D_{n}=\sum_{j=0}^{n} R^{j}D_{n}^{j} ,~~~A_n(x,y) = \sum_{j=0}^{n+1} R^{j} A_{n}^{j}(x,y)  ,
 \end{equation} 
where $D_{n}^{j}$ is the sum of all possible ways to change $j$ of the $n$ columns of $D_n$ from $\bar K$ to $Q$ and $A_{n}^{j}(x,y)$ is the sum of all possible ways to change $j$ of the $n+1$ columns of $A_n(x,y)$ from $\bar K$ to $Q$, namely,
\begin{widetext} 
\begin{eqnarray} 
 \quad D_{n}^{j}=\sum_{1\leq i_{1}<\dots<i_{j}\leq n}
\int\prod_{i=1}^{n}\dd x_{i}
\left|\begin{matrix}
\bar K(x_{1},x_{1}) & \cdots & Q(x_{1},x_{i_{1}}) & \cdots & Q(x_{1},x_{i_{j}}) & \cdots & \bar K(x_{1},x_{n})\\
 & \cdots& & \cdots &  &\vdots  \\
\bar K(x_{n},x_{1}) &\cdots & Q(x_{n},x_{i_{1}}) & \cdots &Q(x_{n},x_{i_{j}}) & \cdots & \bar K(x_{n},x_{n})
\end{matrix}\right|     ,
\end{eqnarray}  
\begin{eqnarray} 
 \quad A_{n}^{j}(x,y)= \!\!\sum_{1\leq i_{1}<\dots<i_{j}\leq n} \!
\int\!\prod_{i=1}^{n}\dd x_{i}
\left|\begin{matrix}
\bar K(x,y) & \bar K(x,x_1) & \cdots & Q(x,x_{i_{1}}) & \cdots & Q(x,x_{i_{j}}) & \cdots & \bar K(x,x_{n})\\
\bar K(x_1,y) & \bar K(x_{1},x_{1}) & \cdots & Q(x_{1},x_{i_{1}}) & \cdots & Q(x_{1},x_{i_{j}}) & \cdots & \bar K(x_{1},x_{n})\\
 & \cdots& & \cdots &  & &\vdots  \\
\bar K(x_n,y) & \bar K(x_{n},x_{1}) &\cdots & Q(x_{n},x_{i_{1}}) & \cdots &Q(x_{n},x_{i_{j}}) & \cdots & \bar K(x_{n},x_{n})
\end{matrix}\right|  .
\end{eqnarray} 
\end{widetext} 
Substituting all these into \eref{fredsol} and expanding perturbatively in terms of $R$, we get the waveform  of the echoes. Explicitly, if we define
\begin{equation}
a_{j}=\sum_{n=j-1}^{\infty}\frac{(-\lambda)^{n}}{n!}A_{n}^{j}(x,y)   ,   \quad b_{j}=\sum_{n=j}^{\infty}\frac{(-\lambda)^{n}}{n!}D_{n}^{j}  ,
\end{equation}
($A_{-1}^{0}(x,y)$ being defined to be 0 for convenience), we can re-write
\begin{align}
\label{general}
\frac{\sum_{n=0}^{\infty}\frac{(-\lambda)^{n}}{n!} A_{n}(x,y) }{\sum_{n=0}^{\infty}\frac{(-\lambda)^{n}}{n!}D_{n}}
=
\frac{\sum_{j=0}^{\infty}a_{j} R^{j}}{\sum_{j=0}^{\infty}b_{j} R^{j}}=\sum_{j=0}^{\infty} c_{j} R^{j}  ,
\end{align} 
where $c_{j}$ can be expressed by $a_{j}$ and $b_{j}$ as:
\begin{equation}
\label{cjexpand}
c_{j}=(-1)^{j}\frac{a_{0}}{b_{0}}\left|\begin{matrix}
\frac{b_{1}}{b_{0}}-\frac{a_{1}}{a_{0}} & 1 & 0 & 0  &\cdots & 0\\
\frac{b_{2}}{b_{0}}-\frac{a_{2}}{a_{0}} & \frac{b_{1}}{b_{0}} & 1  &0 & \cdots &0 \\
\frac{b_{3}}{b_{0}}-\frac{a_{3}}{a_{0}} & \frac{b_{2}}{b_{0}} & \frac{b_{1}}{b_{0}} &1  &\cdots & 0 \\
\cdots  & \cdots&   \cdots     &   \cdots&  \cdots& 1 \\
\frac{b_{j}}{b_{0}}-\frac{a_{j}}{a_{0}} & \frac{b_{j-1}}{b_{0}} & \frac{b_{j-2}}{b_{0}} &\cdots  & \cdots & \frac{b_{1}}{b_{0}}
\end{matrix}
\right|  ,
\end{equation}
and $c_{0}$ is defined as $c_{0}=a_0/b_{0}$. 

Again the integrals of the determinants can be systematically computed by the diagrammatical method outlined in Section~\ref{feynman}. 

In summary, the $n$-th echo can be written as
\begin{equation}
R^n \phi_n(x)=\lambda\int \dd y f(y) c_n(x,y,\omega,\lambda) R(\omega)^n  ,
\end{equation}
where $c_n$ is defined in \eref{cjexpand}.

\section{Diagrammatical representation}
\label{feynman}

In this section, we will show how $\Delta_K(x,y;\lambda)$ and $\Delta(\lambda)$ can be expanded with terms defined in Eq.~(\ref{Andef}) and Eq.~(\ref{Dndef}) respectively, which can be elegantly done with a diagrammatic scheme. We will see that echoes can also be represented within this scheme.
  
This scheme is based on the realization that $K(x_{i},x_{j})$ can be viewed as a 2-point ``vertex'' while an integration $\int\d x$ can viewed as a ``propagator''. In addition, for each diagram with an odd number of loops, we need to assign a minus sign, which is simply due to the antisymmetric nature of the determinant. Furthermore, by \eref{Kdef1} and \eref{Bdef1}, we can also equate the integration of $K$ to the trace of matrix $B$, so we can directly translate the diagrams to traces of matrix $B$. It might be instructive to give a few examples:
\begin{align}
&K(x,y)=\propagaheng ~ ,~~~
\int\dd x K(x,x)={\rm Tr}(B)=\oneloop  ~, \\
&\int\dd x K(x_{1},x)K(x,x_{2})=
\propagatwo  ~ ,\\
&~~~~\int\dd x_{1}\dd x_{2} \dd x_{3} K(x_{1},x_{2})K(x_{2},x_{1})K(x_{3},x_{3})\\
&={\rm Tr}(B^{2}){\rm Tr}(B)=
\twoloop\indent \oneloop   .
\end{align}
In this language, the $n$-dimensional determinant $D_{n}$ can be denoted as a sum of all possible allowed diagrams, each of which contains $n$ vertices connected $n$ propagators without loose ends (ie, each diagram being a multiplication of loops). For example, for $D_4$, we have 
\begin{widetext}
\begin{align}
D_{4}&=\int\prod_{i=1}^{4}\dd x_{i}
\left|\begin{matrix}
K(x_{1},x_{1}) & K(x_{1},x_{2}) & K(x_{1},x_{3}) &  K(x_{1}, x_{k})\\
K(x_{2},x_{1}) & K(x_{2},x_{2}) & K(x_{2},x_{2}) &  K(x_{2},x_{n})\\
K(x_{3},x_{1}) & K(x_{3},x_{2}) & K(x_{3},x_{3}) &  K(x_{3},x_{4})\\
K(x_{4},x_{1}) & K(x_{4},x_{4}) & K(x_{4},x_{3}) &  K(x_{4},x_{4})
\end{matrix}\right|\\
&=(
\oneloop)^{4} 
-
6\indent 
\twoloop
\indent (
\oneloop
)^{2}
+3 \indent
\twoloop
+8 \indent 
\threeloop \indent
\oneloop
-
3! \indent
\floop\\
&={\rm Tr}(B)^{4}- 6{\rm Tr}(B^{2}){\rm Tr}(B)^{2}+3{\rm Tr}(B^{2})^{2}+8 {\rm Tr}(B^{3}){\rm Tr}(B)-6{\rm Tr}(B^{4})   .
\label{D4expansion}   
\end{align}
\end{widetext}
This of course can be verified by the well-known result for the determinant. Since $\Delta(\lambda)$ is just $\det(I-\lambda B)$, we have
\begin{align}\label{formula1}
\Delta(\lambda)=\det(I-\lambda B) =\exp(-\sum_{n=1}^{\infty}\frac{\lambda^{n}}{n}{\rm Tr}(B^{n}))   ,
\end{align}
and then $D_{4}$ is simply $\d^4\Delta(\lambda)/\d \lambda^4|_{\lambda=0}$, which agrees with \eref{D4expansion}. Note that a simple consistent check is that for each $D_n$ expansion the sum of coefficients should add up to zero.

To compute the waveform in a finite interval, we use the approximation $K_{N}(x,y)=\sum_{i=1}^{N}\alpha_{i}(x)\beta_{i}(y)$. In this case, we can truncate the $\lambda$ expansion of $\Delta(\lambda)$ at order $N$, as higher order $D_n$'s vanish identically.

For $\Delta_K(x,y;\lambda)$, the diagrams are not closed loops and extra ``cutting rules'' are needed, as it contains ``external line" $x$ and $y$. To find these rules, we note that
\begin{equation}
D_{n}=\int\dd x A_{n-1}(x,x)   .
\end{equation}
Conversely, $A_{n}(x,y)$ can be obtained by cutting out a propagator of $D_{n+1}$. (Of course, $A_{0}(x,y)$ can not be obtained by cutting, but it is simply $A_{0}(x,y)=K(x,y)$.) When doing the cutting, it is important to remember that we need to apply the Leibniz rule to cut each part of the diagram (eg, a diagram may have 3 parts, ie, 3 loops multiplied together), and for each diagram we cut exactly once. For example, by cutting a line in the diagrams of $D_4$, we can get
\begin{align}
A_{3}(x,y)&=
\propagaheng
\indent
(\oneloop)^{3}
-3 \propagatwo\indent
(\oneloop)^{2}
\nn
&~~~~
-3
\propagaheng \indent\oneloop\indent\twoloop 
+3\propagatwo\indent\twoloop  
\nn
&~~~~
+ 6 \propagathree\indent\oneloop+2\propagaheng\indent\threeloop 
\nn
&~~~~ -6 \propagafour 
\\
&=K(x,y){\rm Tr}(B)^{3}-3({\rm Tr}(B))^{2}\alpha\cdot B\cdot\beta
\nn
&~~~~ -3K(x,y){\rm Tr}(B){\rm Tr}(B^{2})+3\alpha\cdot K\cdot\beta {\rm Tr}(B^{2})
\nn
&~~~~+4\alpha\cdot B^{2}\cdot\beta {\rm Tr}(B)+2K(x,y){\rm Tr}(B^{3})
\nn
&~~~~ -6\alpha\cdot B^{3} \cdot\beta   ,
\end{align}
where we have used the shorthand notation $\alpha\cdot B^{n}\cdot\beta = \sum_{i,j} \alpha_{i}(x)(B^{n})_{ij} \beta_{j}(y)$. The coefficients in front of the diagrams are obtained by
\bal
& a_i^{m_i}a_j^{m_j} \xrightarrow{\rm cut} \\
& \frac{im_i}{im_i+jm_j}b_ia_i^{m_i-1}a_j^{m_j}+\frac{jm_j}{im_i+jm_j}b_j a_i^{m_i}a_j^{m_j-1} , \nonumber
\eal
where $a_i$ denotes a loop with $i$ vertices and $b_i$ denotes a line with $i$ vertices. Again, for each $A_n$ expansion, the sum of coefficients should add up to zero, as the cutting does not change the total number of diagrams.

Therefore, the waveform can be represented as
\bal
&\phi(x) =f(x)-\lambda\! \int\!\!\dd y\f{({\text{\small diagrams with 1 cut}})}{({\text{\small closed diagrams}})} f(y)
\\&=f(x) -\lambda\int\dd y\frac{\text{cut}{\rm ~exp} {\small (-\lambda\oneloop-\frac{\lambda^{2}}{2}\twoloop-\cdots) } }{{\rm exp}(-\lambda\oneloop-\frac{\lambda^{2}}{2}\twoloop-\cdots)} f(y)  .
\eal
Note that the cutting operation does not commute with the exponential map, so we need to compute the exponential first and then perform the cutting.

Having established the general diagrammatical formalism to compute waveforms, it is straightforward to extract the echoes form it. Note that, for scenarios with echoes, the kernel function may be split into two parts
\be
K(x,y) = \bar K(x,y)+R(\omega) Q(x,y)  .
\ee
This suggests that we can add an extra 2-point vertex representing $ Q(x,y)$:
\be
Q(x,y) = \onedotline  .
\ee
(In this notation, of course, the 2-point circle vertex represent $\bar K(x,y)$ or not $K(x,y)$.) 
Therefore, all the loop diagrams in the general formalism should be replaced with
\begin{align}
\oneloop &\rightarrow \oneloop + R(\omega)\onedotloopone   , \\
\twoloop &\rightarrow \twoloop + 2R(\omega)\onedotlooptwo + R^{2}(\omega) \twodotlooptwo  , \\
&\cdots  .
\end{align}
$\Delta(\lambda)$ then is expanded as
\begin{equation}
\Delta(\lambda)= \sum_{n=0}^{\infty}\frac{(-\lambda)^{n}}{n!}D_{n}
\end{equation}
with
\begin{equation}
\begin{split}
D_{0}&=1 ,  \\
D_{1}&=\int \dd x_{1}(\bar K_{11}+R(\omega) Q_{11})   
\nn
&=\oneloop+R(\omega)\onedotloopone ,  \\
D_{2}&=\int\dd x_{1}\d x_{2}\Bigg[
\left|\begin{matrix}
\bar K_{11} & \bar K_{12}\\
\bar K_{21} & \bar K_{22}
\end{matrix} 
\right|
\nn
& ~~~~~~ +R(\omega) \left(
\left|\begin{matrix}
\bar K_{11} & Q_{12}\\
\bar K_{21} & Q_{22}
\end{matrix} \right|
+
\left|\begin{matrix}
Q_{11} & \bar K_{12}\\
Q_{21} & \bar K_{22}
\end{matrix} \right|
\right)
\nn
&~~~~~~ +R(\omega)^{2} \left|\begin{matrix}
Q_{11} & Q_{12}\\
Q_{21} & Q_{22}
\end{matrix} \right|
\Bigg]\\
&= (\oneloop)^{2}-\twoloop
\nn
&~~~ + 2 R(\oneloop\indent\onedotloopone-\onedotlooptwo)
\nn
&~~~ +R^{2}((\onedotloopone)^{2}-\twodotlooptwo)   ,
\\
&\cdots  , \nonumber
\end{split}
\end{equation}
where we have used the short note $\bar K_{ij}$ and $Q_{ij}$ for $\bar K(x_{i},x_{j})$ and $Q(x_{i},x_{j})$. The cutting rules also apply, so we have
\be
\sum_{n=0}^{\infty}\frac{(-\lambda)^{n}}{n!}A_{n}(x,y)   
\ee
with
\begin{equation}
\begin{split}
A_{0}(x,y)&=\text{cut}D_{1}=\oneline+ R(\omega) \onedotline  , \\
A_{1}(x,y)&=\text{cut}D_{2}= \oneline\quad\oneloop-\twoline
\nn
&+R(\omega)(\oneline\onedotloopone+\onedotline\quad\oneloop-2\quad\onedotlinetwo)
\nn
&+R^{2}(\omega)(\onedotloopone\onedotline-\twodotlinetwo)  ,
\\
&\cdots .  \nonumber
\end{split}
\end{equation}
Therefore, to calculate the $n$-th echo signal, we should collect all the diagrams with $n$ 2-point solid vertices.

Let us summarize  the diagrammatical rules that are needed in order to compute the waveform and echoes:
\begin{itemize}
\item[$\bullet$] $\bar K(x_{i},x_{j})$ is represented as a (2-point) circle vertex;

\item[$\bullet$] $Q(x_i,x_j)$ is represented as a (2-point) solid vertex;

\item[$\bullet$] Diagrams with $n$ solid vertices contribute to the $n$-th echo waveform;

\item[$\bullet$] $\int\d x$ is represented as a ``propagator'';

\item[$\bullet$] For each diagram with an odd number of loops, assign a minus sign;

\item[$\bullet$] $D_n$ includes all loop diagrams with $n$ vertices;

\item[$\bullet$] $A_n$ is obtained by cutting open one of the loops in $D_{n+1}$.
\end{itemize}

     \section{Numerical results in the Fredholm approach}
     \label{numerical}
     
We have shown that the Fredholm approach can be used perturbatively to process individual echo waveforms and for separable cases to conveniently compute the exact waveform analytically. In this section, we shall demonstrate how to compute the waveform for generic (non-separable) cases. 

To numerically compute the full waveform, the simplest way is simply to discretize the Fredholm integral equation \eref{fredhomeq} in the finite interval $[a,b]$ as 
\begin{align}
\label{mtx}
f_i(\omega)&= \sum_{j=1}^N  B_{ij}(\omega)\phi_j(\omega), ~~~(i=1,2,...,N)   ,
\\
    &=\phi_i(\omega)-\frac{b-a}{3N}\sum_{j=1}^N\mathcal{K}_{ij}(\omega)\phi_j(\omega) ,
\end{align}
where  $N$ is an even number and we have defined
\begin{align}
f_i(\omega) &=\int G(\omega,x_i,y)\mathcal{I}(\omega,y)\dd y  ,
\\
x_i&=a+\frac{i}{N}(b-a)   ,
\\
B_{ij}(\omega) &=\delta_{ij}-\frac{b-a}{3N}\mathcal{K}_{ij}(\omega)  ,
\\
\label{KijG}
\mathcal{K}_{ij}(\omega)&=\eta_j G(\omega,x_i,x_j)V(x_j)   ,
\\
\eta_j& =\left\{
\begin{array}{rcl}
1       &      & {j=0,N}\\
4     &      & {j=1, 3, ..., N-1}\\
2     &      & {j=2, 4, ..., N-2}
\end{array} \right.   .
\end{align}
Here we use the Simpson's rule for numerical integration, as coded in the numerical factor $\eta_j$. Thus the waveform computation reduces to a linear algebraic problem, for which there exist a vast collection of fast numerical libraries to facilitate efficient and precision computations. To get the waveform in the time domain, we evaluate $\phi(\omega,x)$ for a large number of $\omega$'s and then we perform the inverse Laplace transformation (see \eref{invLap}).

As an aside, we want to point out a simple application of the Fredholm formalism to the anti-scattering problem --- extract the form of the potential via the waveform. With the method we introduced here, it is straightforward to re-construct the potential from the waveform in the frequency domain. A simple reformulation of \eref{mtx} gives
\begin{equation}
\frac{3N}{b-a}(\phi_i-f_i)=\sum_{j=1}^N\eta_jG(x_i,x_j)\phi_jV(x_j)   .
\end{equation}
Defining matrix $C_{ij}=\eta_jG(x_i,x_j)\phi_j$ and vector $\tilde{\phi}_i=\frac{3N}{b-a}(\phi_i-f_i)$, the potential vector is then simply given by
\begin{equation}
V(x_i)=(C^{-1})_{ij}\tilde{\phi}(x_j)  ,
\end{equation}
where $C^{-1}$ is the inverse of matrix $C$. This is of course easy to solve. In the gravitational wave setting, this is not useful since we can only observe the waveform far away from the source. However, it may find its application in a laboratory setting for a system satisfying the Schroedinger equation.

In the following, we shall compute the waveforms of a few idealized models to demonstrate the efficiency and accuracy of our method. For all these simulations, the initial conditions we use are
\begin{equation}
\Phi(0,x)=e^{-x^2/2}, ~~~~\dot{\Phi}(0,x)=0   .
\end{equation}
The models we consider are tabulated as follows,
\bc
\begin{tabular}{p{3em}p{8em}p{9em}p{2.7em}}
\hline
Model & Potential $V(x)$  & Left Boundary & Figure\\
\hline
(a) & $V_{\rm gr}(x)$  &open & Fig.~\ref{fig:vgr}\\
(b) & $V_{\rm gr}(x)+V_{\rm rec1}(x)$  & open &Fig.~\ref{fig:3pot}\\
(c) & $V_{\rm gr}(x)$ & Dirichlet ${\footnotesize x=-20}$ &Fig.~\ref{fig:mir}\\
(d) & $V_{\rm gr}(x)$  & Neumann ${\footnotesize x=-20}$  &Fig.~\ref{fig:mir}\\
(e) & $V_{\rm gr}(x)+V_{\rm rec2}(x)$ &open &Fig.~\ref{fig:3pot}\\
\hline
\end{tabular}
\ec
with
\begin{align}
V_{\rm gr}(x)&= \(1-\dfrac{1}{r(x)}\)\(\dfrac{l(l+1)}{r^{2}(x)}+\dfrac{(1-s^{2})}{r^{3}(x)}\)  ,
\\
V_{\rm rec1}(x)&=\left\{
\begin{array}{rcl}
2       &      & {-50<x<-49}\\
0     &      & {\rm otherwise}
\end{array} \right.  ,
\\
V_{\rm rec2}(x)&=\left\{
\begin{array}{rcl}
2       &      & {-80<x<-79}\\
2       &      & {-50<x<-49}\\
0     &      & {\rm otherwise}
\end{array} \right.  .
\end{align}
where we have set $2GM=1$ and again $r(x)$ is obtained by inverting the relation $x=r+\ln({r-1})$ and in this section we shall set $l=2$ and $s=2$.

\begin{figure}[h]\centering
\includegraphics[width=.45\textwidth]{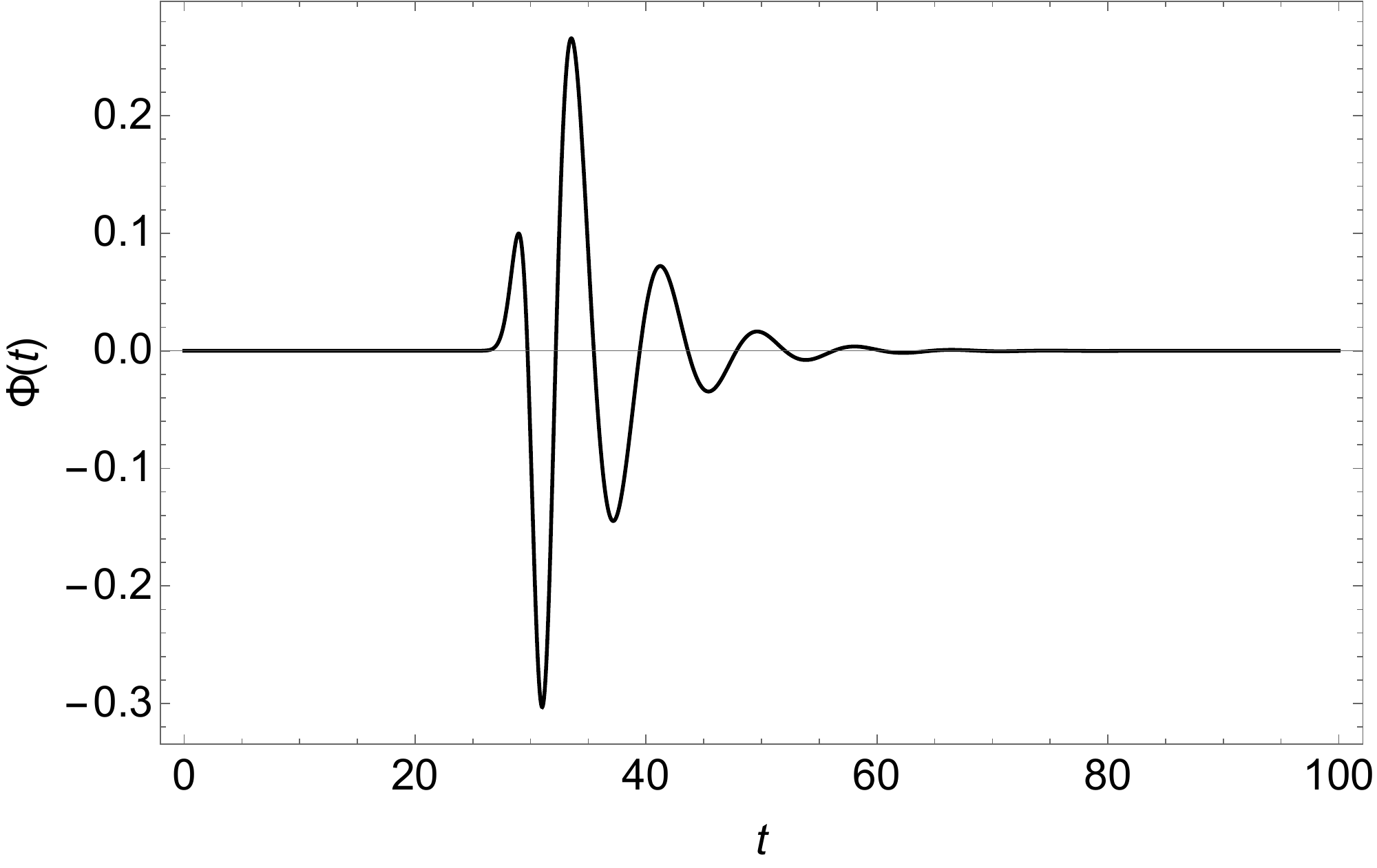}
\includegraphics[width=.45\textwidth]{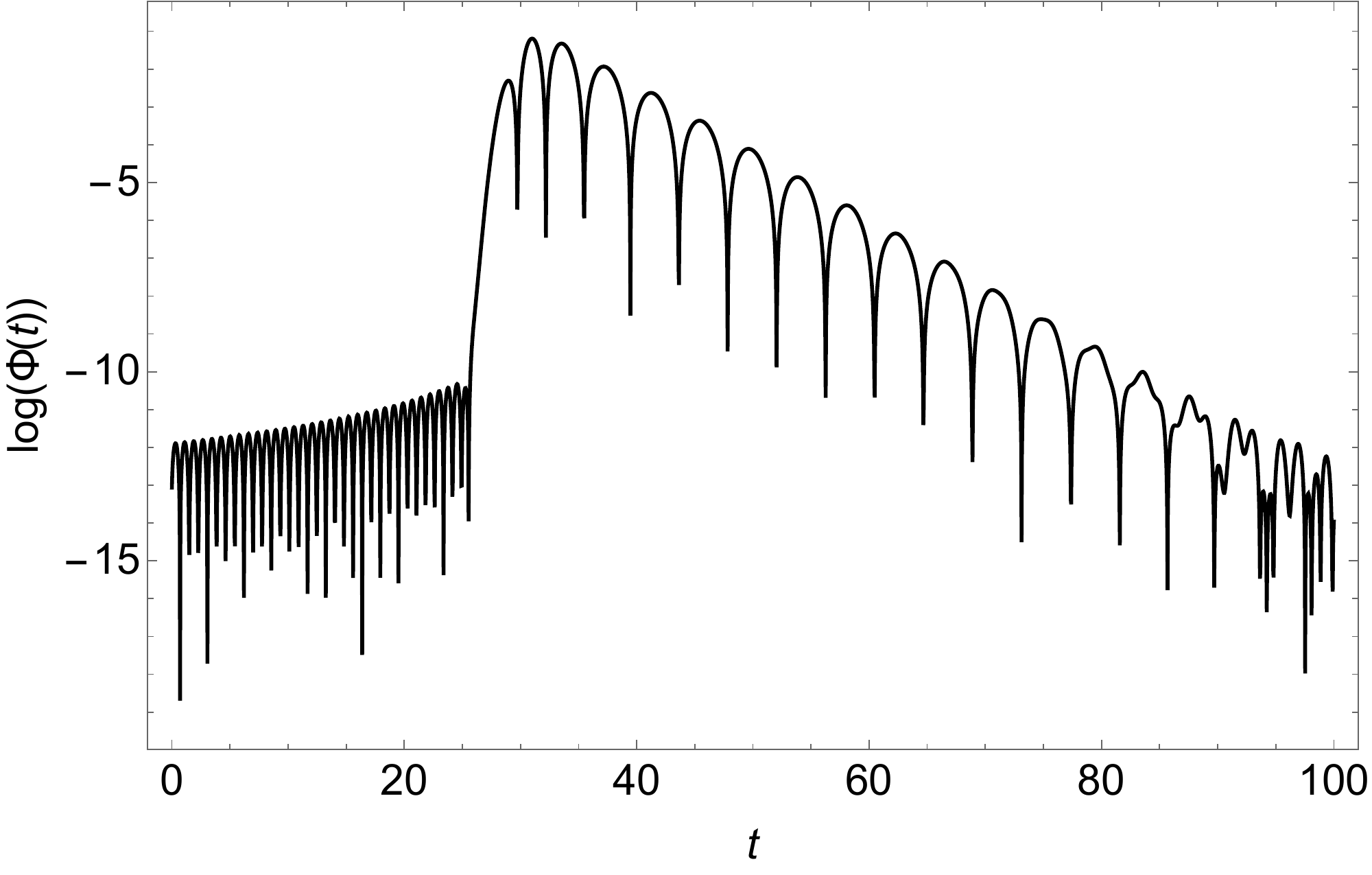}
\caption{Waveform generated by Model (a), ie, the Schwarzschild case. We choose the numerical parameters $a=-8$, $b=113$, $N=484$ and observe the waveform at $x=30$. The initial conditions are $\Phi(0,x)=e^{-x^2/2}, \dot{\Phi}(0,x)=0$ and we set $l=2$ and $s=2$ for $V_{\rm gr}(x)$, the same for all the figures in this section.
}
\label{fig:vgr}
\end{figure}

Fig.~\ref{fig:vgr} is just the ringdown waveform generated by the Schwarzschild effective potential $V_{\rm gr}(l=2,s=2)$, observed at $x=30$. As we shall see shortly, even for a relatively narrow interval and small $N$, our method already generates very accurate waveforms. 

\begin{figure}[h]
\includegraphics[width=.45\textwidth]{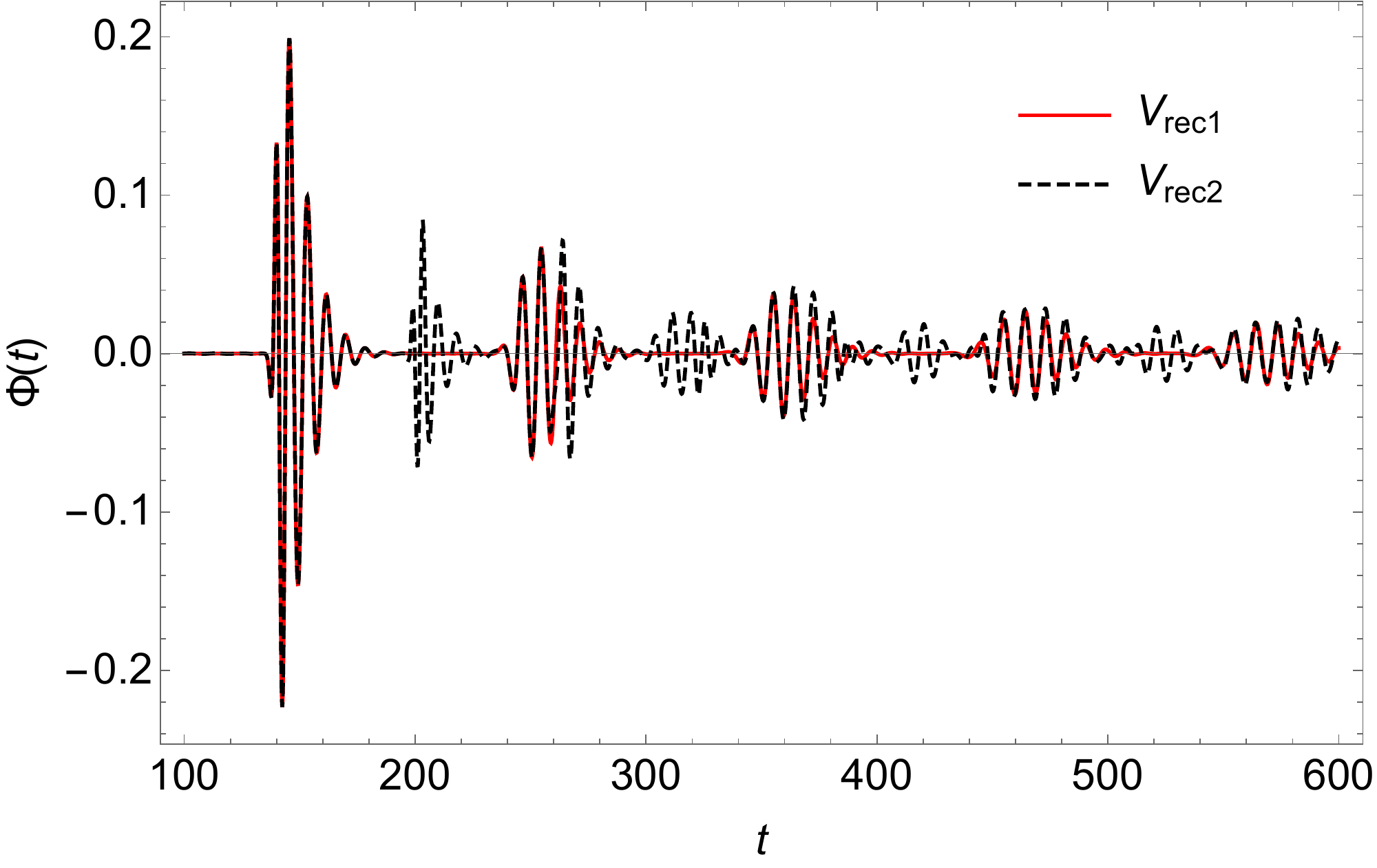}
\caption{Waveforms generated by Model (b) and (e). The red solid line is the waveform generated by $V(x)=V_{\rm gr}(x)+V_{\rm rec1}(x)$, with parameters $a=-80$, $b=83$ and $N=652$, while the black dashed line is the waveform generated by $V(x)=V_{\rm gr}(x)+V_{\rm rec2}(x)$ with parameters $a=-80$, $b=83$, $N=652$. We observe the waveform at $x=40$. Model (e) contains all the echoes of Model (b) and also has extra smaller echoes in between the bigger echoes.
}
\label{fig:3pot}
\end{figure}

Fig.~\ref{fig:3pot} is a comparison between the two waveforms generated by Model (b) ($V(x)=V_{\rm gr}(x)+V_{\rm rec1}(x)$) and Model (e) ($V(x)=V_{\rm gr}(x)+V_{\rm rec2}(x)$), observed at $x=40$. Model (e) has an extra potential barrier to the left of the potential peak of Model (b). We see that Model (e) contains all the echoes of Model (b) and also has extra smaller echoes in between the bigger echoes of Model (b), as anticipated from the intuition of a propagating wave package transmitting and reflected between the potential barriers before reaching the observer to the right of the Schwarzschild potential barrier. In particular, the bigger echoes are due to the waves bouncing back and forward between the Schwarzschild potential peak and the rectangular potential at $x=-50$, while the smaller echoes come from waves bouncing back and forward between the rectangular potentials at $x=-50$ and $x=-80$. The latter ones are relatively smaller because they have to transmit through an extra potential barrier.

\begin{figure}[h]
\includegraphics[width=.45\textwidth]{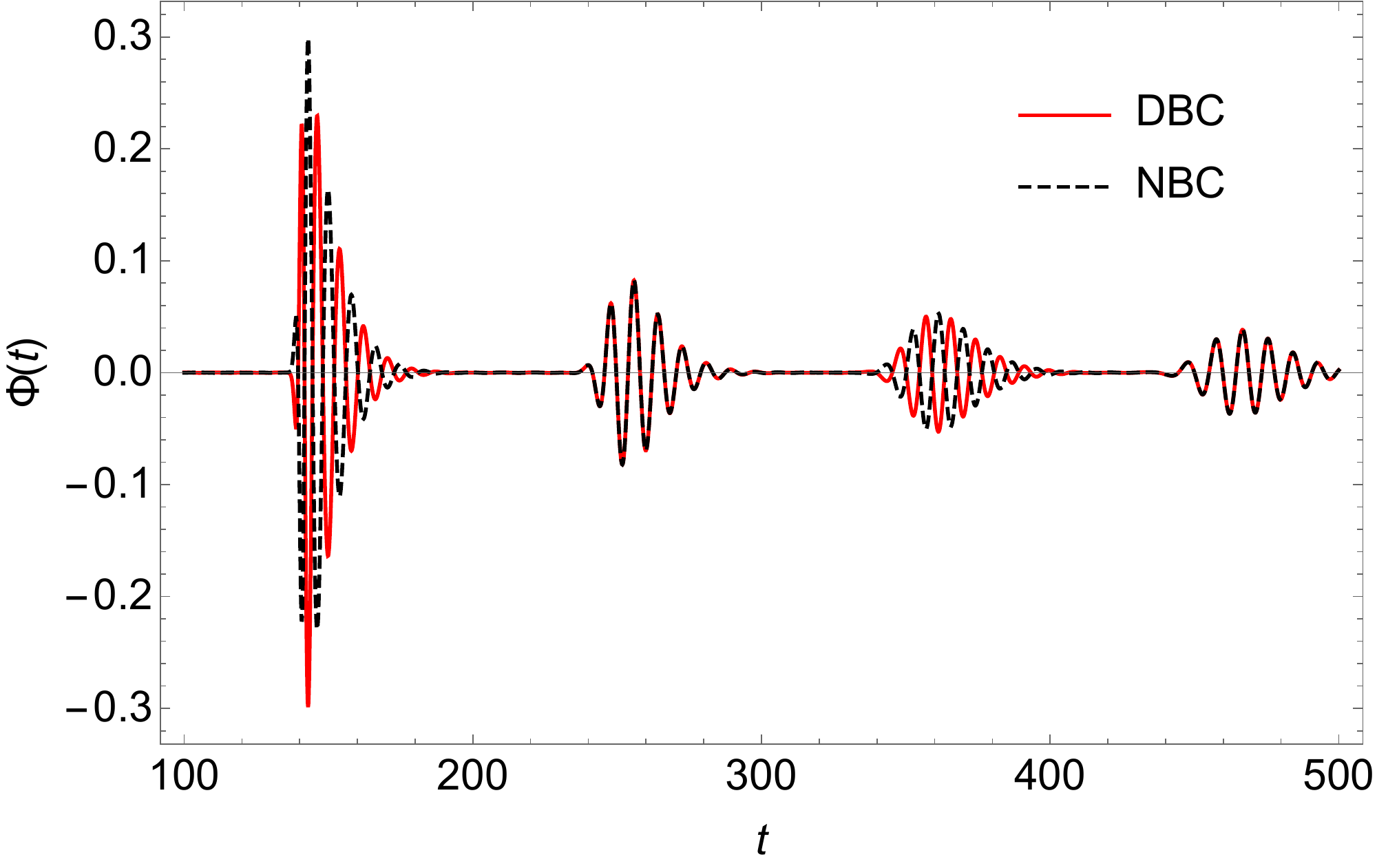}
\caption{Waveforms generated by Model (c) and (d).  We have chosen the parameters to be $a=-52$, $b=83$, $N=540$ and observe the waveform at $x=40$. The reflecting ``mirror'' is at $x=-L=-50$. The red solid line is generated by a ``mirror'' satisfying the Dirichlet boundary condition (DBC) $R(\omega)=-e^{i\omega 2L}$, while the black dashed line is generated by a ``mirror'' satisfying the Neumann boundary condition (NBC) $R(\omega)=e^{i\omega 2L}$.
}
\label{fig:mir}
\end{figure}

When there is a reflecting ``mirror'' at $x=-L$, we should use the Green's function \eref{mirgf} with nonzero $R(w)$. For an illustration, we consider a simple frequency independent reflectivity
\begin{equation}
R(\omega)=-r_fe^{i2\omega  L}   ,
\end{equation}
where $r_f=1$ for the Dirichlet boundary condition and $r_f=-1$ for the Neumann boundary condition. The difference of these two boundary conditions is of course to flip the value of waveform $\phi(t,x)$ for echoes associated with odd powers of $R(\omega)$ (see Fig.~\ref{fig:mir}).

\begin{figure*}[ht]\centering
\includegraphics[width=.45\textwidth]{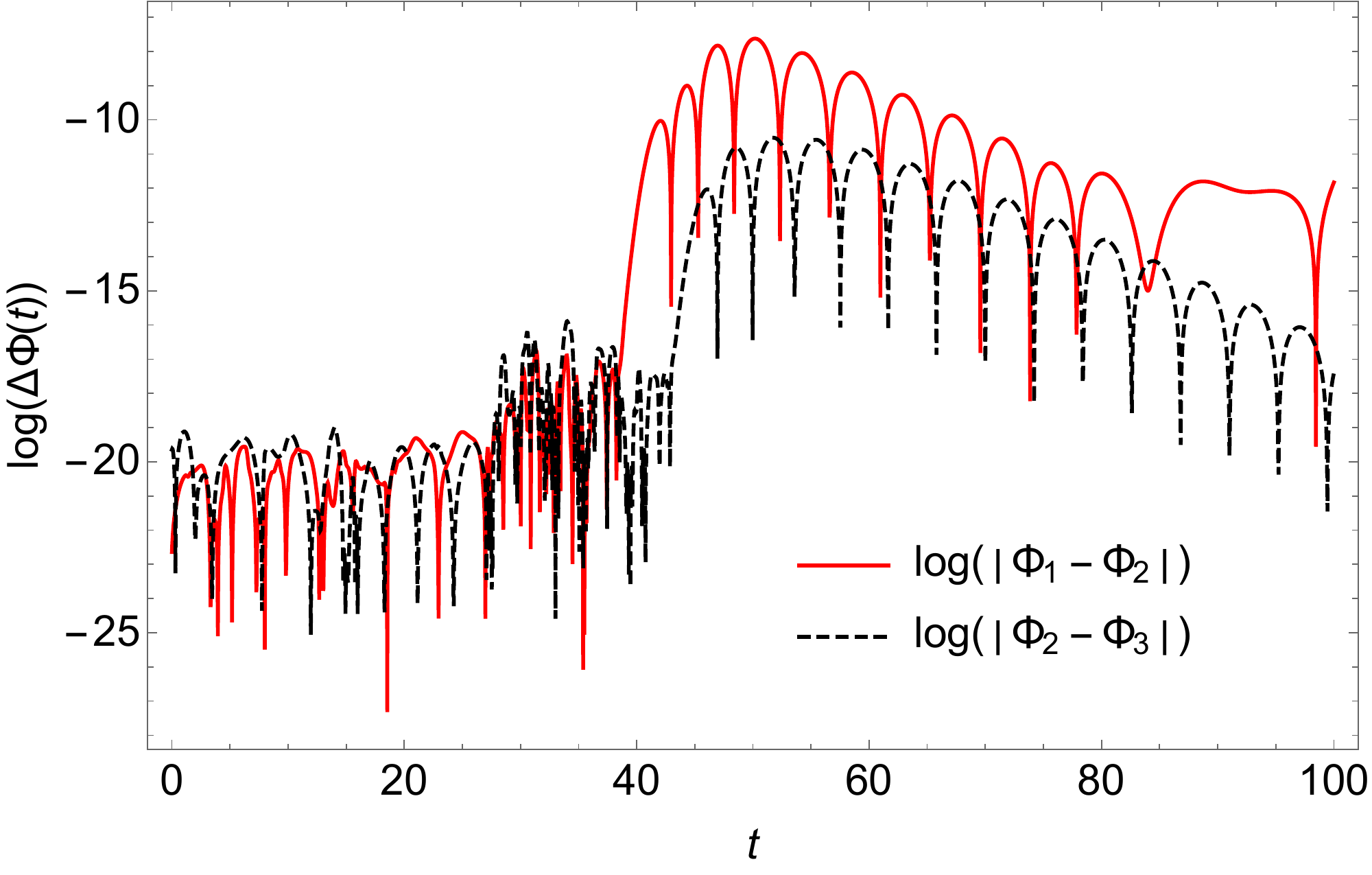}
\includegraphics[width=.45\textwidth]{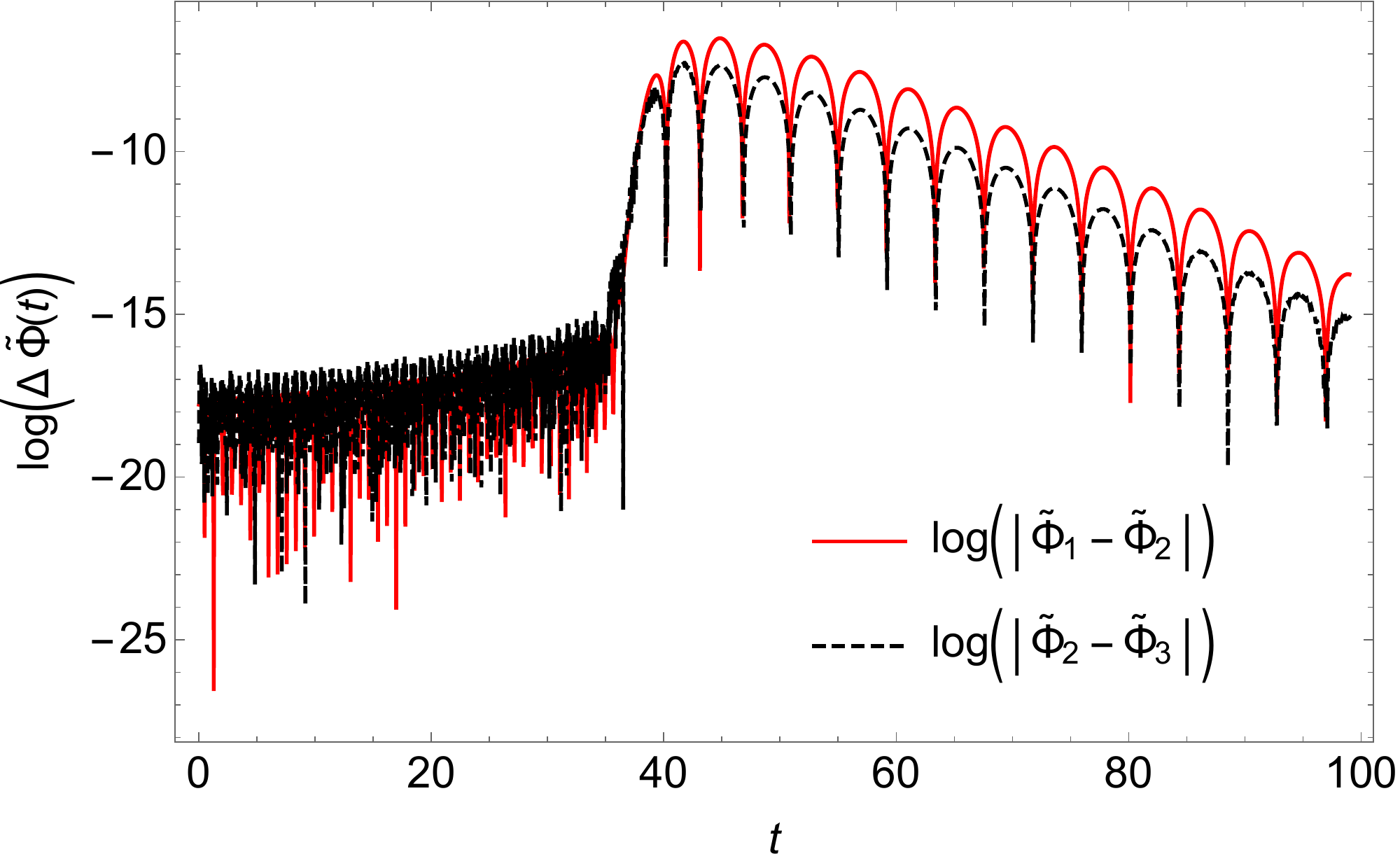}
\caption{The left figure shows the differences between time domian waveforms ($\Phi_1,\Phi_2$ and $\Phi_3$) generated by 3 different intervals in Model (a).  The red solid line represents $\log|\Phi_1-\Phi_2|$ and the black dashed line represents $\log|\Phi_2-\Phi_3|$. The 3 intervals are chosen as: $a=-6$ and $b=38$ for $\Phi_1$, $a=-8$ and $b=113$ for $\Phi_2$ and $a=-10$ and $b=350$ for $\Phi_3$, which essentially cut off the potential at approximately $V_{\rm gr}=0.005, 0.0005$ and $0.00005$ respectively. All of them have $N=4$ grid points per length. The right figure shows the difference between time domain waveforms ($\tilde\Phi_1,\tilde\Phi_2$ and $\tilde\Phi_3$) generated by 3 different grid points in Model (a). The red solid line represents $\log|\tilde{\Phi}_1-\tilde{\Phi}_2|$ and the black dashed line represents $\log|\tilde{\Phi}_2-\tilde{\Phi}_3|$. We choose $\tilde{\Phi}_1$ to have $N=4$ grid points per length, $\tilde{\Phi}_2$ to have $N=8$ grid points per length and $\tilde{\Phi}_3$ to have $N=16$ grid points per length. All of them are calculated in interval $[-7,83]$.}
\label{conv1}
\end{figure*}

In our numerical scheme, we have truncated the potentials for $x<a$ and $x>b$. We previously have shown that this is a consistency scheme by proving the convergence and estimating the errors. Numerically, the effectiveness of this approximation can also be verified by convergence studies, in which we do computations with different ranges of the interval $[a,b]$ and check whether the waveforms converge to a waveform with enlarged intervals. In Fig.~\ref{conv1}, we plot the differences between the 3 waveforms generated by 3 different intervals in Model (a), ie, $V_{\rm gr}$ $(l=2$, $s=2)$. We see that the waveforms converge to the real values very quickly even for relatively small intervals.

Because of the good convergence properties, our scheme is insensitive to the boundary conditions.  We also want to point out that the computation costs do not sensitively rely on the complexity of the effective potential, as we only need to solve a system of linear equations.  

Finally, let us note that the Fredholm approach also presents a way to compute the quasi-normal modes, which are simply the solutions to $\det(B)=0$. For example, the fundamental quasi-normal mode for a rectangular potential
\begin{equation}
V_{\rm rec}(x)=\left\{
\begin{array}{rcl}
16       &      & {0<x<1}\\
0     &      & {\rm otherwise}
\end{array} \right.
\end{equation}
can be analytically calculated and is $4.660-0.710i$, see \cite{Barausse:2014tra}. In our approach, we only need $N=26$ to reach the accuracy of 1\%, ie, $\det(B)=10^{-2}$.

\section{Conclusions}

Apart from the confirmation of the most spectacular natural phenomena themselves, the newly arrival of the gravitational wave astronomy has allowed us to probe the astonishing properties of compact astronomical objects and gravitation itself, particularly in the strong field regime. While initially considered as rather exotic objects, black holes have long established their roles in modern astrophysics. The gravitational wave astronomy ushers in the exciting possibility of detecting potentially more surprising attributes of the black holes or even more exotic compact objects, leading to new discoveries in fundamental physics.

Echoes in the ringdown waveform are messengers of many near-horizon exotic new physics, and thus it is justified to carefully characterize and accurately describe the essential features of the waveforms containing them so as to help build reasonable templates to extract useful information in the observational data. We have proposed a new method to represent the individual echoes as terms of different orders in a Pade type perturbative expansion. Also, we have devised an intuitive diagrammatical scheme involving vertices, propagators and cutting rules to facilitate the perturbative expansion. We have also proven the convergence of the Pade expansion, estimated the errors and numerically demonstrated efficiency of the method. For some separable cases, our approach also allows for an easy computation of the exact non-perturbative wave solution.

~\\    

    \noindent{\bf Acknowledgments}:
We would like to thank Dong-Ming He, Zhi-Peng Li, Yun-Song Piao, Hao-Yu Shang, Ying-Jie Wang and Jun Zhang for helpful discussions. We are especially grateful to Wen-Bo Li for discussions on some mathematical aspects in the paper.  SYZ acknowledges support from the starting grant from University of Science and Technology of China under grant No.~KY2030000089 and is also supported by National Natural Science Foundation of China under grant No.~GG2030040375.\\

~\\

    \bibliography{refs.bib}
    
\end{document}